\documentclass[twocolumn]{aastex61}
\pdfoutput=1 
\usepackage{amsmath}
\usepackage[T1]{fontenc}
\usepackage{url}
\usepackage{graphicx}
\usepackage[caption=false]{subfig}

\usepackage{natbib}
\bibpunct{(}{)}{;}{a}{}{,} 

\begin{document}

\title{TeV gamma-ray observations of the binary neutron star merger GW170817 with H.E.S.S.} 

\AuthorCallLimit=300
\collaboration{(H.E.S.S. Collaboration)}\altaffiliation{corresponding author (contact.hess@hess-experiment.eu)}
\noaffiliation\email{contact.hess@hess-experiment.eu}
\author{H.~Abdalla}\affiliation{Centre for Space Research, North-West University, Potchefstroom 2520, South Africa}
\author{A.~Abramowski}\affiliation{Universit\"at Hamburg, Institut f\"ur Experimentalphysik, Luruper Chaussee 149, D 22761 Hamburg, Germany}
\author{F.~Aharonian}\affiliation{Max-Planck-Institut f\"ur Kernphysik, P.O. Box 103980, D 69029 Heidelberg, Germany}\affiliation{Dublin Institute for Advanced Studies, 31 Fitzwilliam Place, Dublin 2, Ireland}\affiliation{National Academy of Sciences of the Republic of Armenia,  Marshall Baghramian Avenue, 24, 0019 Yerevan, Republic of Armenia}
\author{F.~Ait~Benkhali}\affiliation{Max-Planck-Institut f\"ur Kernphysik, P.O. Box 103980, D 69029 Heidelberg, Germany}
\author{E.O.~Ang\"uner}\affiliation{Instytut Fizyki J\c{a}drowej PAN, ul. Radzikowskiego 152, 31-342 Krak{\'o}w, Poland}
\author{M.~Arakawa}\affiliation{Department of Physics, Rikkyo University, 3-34-1 Nishi-Ikebukuro, Toshima-ku, Tokyo 171-8501, Japan}
\author{M.~Arrieta}\affiliation{LUTH, Observatoire de Paris, PSL Research University, CNRS, Universit\'e Paris Diderot, 5 Place Jules Janssen, 92190 Meudon, France}
\author{P.~Aubert}\affiliation{Laboratoire d'Annecy-le-Vieux de Physique des Particules, Universit\'{e} Savoie Mont-Blanc, CNRS/IN2P3, F-74941 Annecy-le-Vieux, France}
\author{M.~Backes}\affiliation{University of Namibia, Department of Physics, Private Bag 13301, Windhoek, Namibia}
\author{A.~Balzer}\affiliation{GRAPPA, Anton Pannekoek Institute for Astronomy, University of Amsterdam,  Science Park 904, 1098 XH Amsterdam, The Netherlands}
% ...


\begin{abstract}
We search for high-energy gamma-ray emission from the binary neutron star merger GW170817 with the H.E.S.S. Imaging Air Cherenkov Telescopes. The observations presented here have been obtained starting only 5.3h after GW170817. The H.E.S.S. target selection identified regions of high probability to find a counterpart of the gravitational wave event. The first of these regions contained the counterpart SSS17a that has been identified in the optical range several hours after our observations. We can therefore present the first data obtained by a ground-based pointing instrument on this object. A subsequent monitoring campaign with the H.E.S.S. telescopes extended over several days, covering timescales from 0.22 to 5.2 days and energy ranges between $270\,\mathrm{GeV}$ to $8.55\,\mathrm{TeV}$. No significant gamma-ray emission has been found. The derived upper limits on the very-high-energy gamma-ray flux for the first time constrain non-thermal, high-energy emission following the merger of a confirmed binary neutron star system. 
\end{abstract}


\section{Introduction}
Opening the era of gravitational wave (GW) astronomy, the first direct detection of a GW signal from a binary black hole merger was reported by the LIGO-Virgo scientific collaboration (LVC) in September 2015~\citep{abbott2016observation} during the first science run (O1) of the Advanced LIGO interferometers. The second science run O2 started in Fall 2016 with the two LIGO detectors taking data. The Advanced Virgo interferometer joined the observations on August 1, 2017. \\
Marking the beginning of gravitational wave multi-messenger astronomy, a gravitational wave signal compatible with that expected from the merger of a binary neutron star system was detected by the LIGO-Virgo collaboration on August 17, 2017~\citep{LVCGW170817PRL}. The event stands as the first direct detection of gravitational waves coming from a system of this kind. As these events are assumed to be related to Gamma-Ray Bursts (GRBs) or kilonovae~\citep{metzger2012most}, broadband emission of electromagnetic (EM) radiation as well as high-energy neutrinos~\citep{eichler:1989aa} can be expected in addition to the gravitational wave signal. Gamma-ray detections in the GeV-TeV energy range have been argued to depend on the specifics of the progenitors ambient environment density, energy fraction in electrons and magnetic fields, the mergers proximity and the viewing angle with respect to the outflow by~\citep[e.g.][]{zhu2016,takami2014}. An extensive observational campaign covering a very wide range of EM wavelengths from radio to high-energy gamma rays and including high-energy neutrinos was launched immediately after the detection of the GW signal. \\
Gamma-Ray Bursts are short bursts of radiation with prompt emission typically detected in the tens of keV to GeV range~\citep[e.g.][]{zhang2006physical}. The duration of the initial, or prompt, emission follows a bimodal distribution, with the divisional timescale between both types of bursts being around 2 seconds~\citep{kouveliotou1993identification}. This distinction  enables their classification into \textit{short} GRB (sGRB) and \textit{long} GRB (lGRB). These classes have been linked to the association of the events with different cosmic progenitors. The long GRBs are usually associated with the core-collapse of massive stars~\citep{woosley2007pulsational, piran2017relativistic}, while the coalescence of a compact binary system, being either a neutron star-neutron star (NS-NS) or a neutron star-black hole, are thought to be the cosmic progenitors of short GRBs that have a hard spectrum~\citep{piran1994gamma, metzger2012most}. In these systems, the orbit of the binary system steadily decays as a result of the energy loss through continuous gravitational wave emission, resulting in the objects spiraling inwards at an increasing rate. At the final phase of the process, a characteristic burst of gravitation radiation is emitted, a prime signal for current GW instruments. In addition, the GW radiation is expected to be accompanied by both thermal and non-thermal emission in the form of EM radiation. Depending on the mass losses during the cataclysmic event, the residual compact object left afterwards will be a black hole or a neutron star. However, due to strong absorption at early times and the beaming effects associated with relativistic outflows of the post-merger ejecta, the EM radiation from the inner engine 
may be shrouded from view. Therefore, a gravitational wave observation of such an event, provides substantial new information to characterize the progenitor system and the phenomena leading to the explosive merger. 

One of the leading theoretical frameworks describing the phenomenology of GRBs is the so-called fireball model ~\citep{meszaros1993relativistic, piran1999gamma}. After the formation of a new compact object, the central engine releases a huge amount of energy over a short time and within a small volume, giving rise to relativistic outflows and shocks. This {\it fireball} is considered to consist of an optically thick electron-positron/photon plasma expanding with relativistic velocities.

Additionally, another class of EM transient counterpart to mergers of binary neutron stars has been proposed. These events are called {\it Macronovae}~\citep{Kulkarni2005} or {\it Kilonovae}~\citep{Metzger2010}, and their energy output lies between the novae and supernovae energy scales. Kilonovae produce delayed optical, UV and infrared radiation on timescales of  a few days, heated by the radioactive decay of $r$-process elements in the ejecta itself, or in the interaction of the ejecta with the interstellar medium~\citep[see e.g.][for a GRB-Kilonova association]{Tanvir2013}. The ejecta in a kilonova are believed to have a mass of $\sim$10$^{-2}M_\odot$ and are moving at mildly-relativistic velocities of 0.1-0.2$c$. Kilonovae produce rather isotropic emission that typically lasts for days after the merger event~\citep[see][and references therein]{Baiotti2017}.

Clear evidence for a non-thermal emission from sGRBs has been found in the GeV energy range for only a handful of cases \citep{Ackermann2013}, with maximum observed photon energies exceeding 30\,GeV \citep{Ackermann2010}. In principle, this gamma-ray emission may be produced via energy losses from particles accelerated at shocks present in the outflow or formed when the ejecta propagate through the interstellar medium. Given the ejected mass and ejecta velocities considered, this situation is reminiscent of a supernova remnant where diffusive shock acceleration would proceed in the non-relativistic to trans-relativistic regime~\citep{Ellison2013}. With its superior sensitivity on short timescales above 50\,GeV, relative to the Fermi-LAT instrument, the High Energy Stereoscopic System (H.E.S.S.\,\,II) is uniquely suited to probe the level of non-thermal emission produced by high energy transient events \citep{HESS_transient_sens}.

High energy observations of non-thermal emission in the GeV-TeV energy range thus provide an effective probe of non-thermal emission from both {\it fireball} and {\it kilonovae} classes of events. 
With the significant sensitivity achieved in this energy range, a detection of a cut-off in the spectral energy distribution of the emission is possible, which would provide hints on the environment of the cataclysmic event~\citep{ackermann2011detection} and allow one to estimate the kinematic velocities of the outflow in which the $\gamma$-ray emission was produced.
Moreover, the comparison between the highest and the low-energy photons from the same source can provide constraints on theories of Lorentz invariance violation~\citep{biesiada2009lorentz}.\\

This article is organized as follows. In Sec.~\ref{Section:NSM}, the gravitational wave event, the subsequent EM follow-up campaign and the H.E.S.S. follow-up effort is discussed. Section~\ref{Section:Data} describes the data and their analysis and in Section~\ref{Section:Results} the results are described. Discussion and conclusion are presented in Section~\ref{Section:Summary}. 

\section{Neutron star merger follow-up}
\label{Section:NSM}
\subsection{Electromagnetic follow-up of GW alerts}
In preparation of the physics data taking of the advanced LIGO and Virgo interferometers, agreements with an extensive group of observatories interested in performing follow-up observations across the EM spectrum and using high-energy neutrinos have been set up by the LIGO-Virgo collaboration. H.E.S.S. became a member of this group in early 2015 and the follow-up of gravitational wave alerts has been prepared (see Sec.~\ref{sec:HESSscheduling} for details).

To rapidly alert the multiwavelength (MWL) follow-up community, the LIGO-Virgo collaboration has developed different low-latency pipelines searching for transient signals from compact binary mergers. The fastest pipeline is BAYESTAR~\citep{singer2016rapid}. It is complemented by the LALInference algorithm, which is scanning a larger parameter space and marginalizing over calibration uncertainties and is thus providing a more robust estimate of the important event parameters~\citep{veitch2015parameter}. 

\begin{figure*}[!th]
  \resizebox{\hsize}{!}{
  \includegraphics[width=0.95\textwidth]{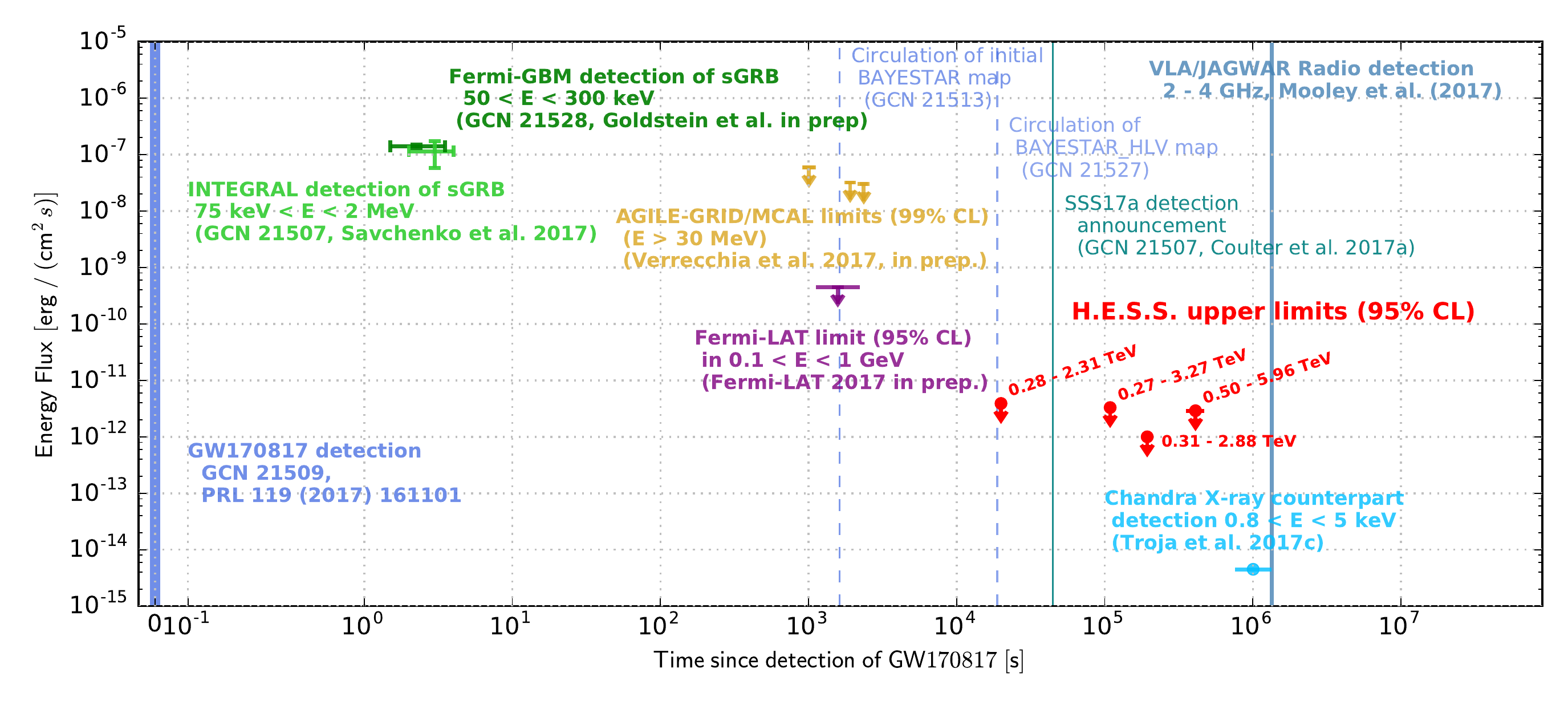}%
  }
\caption{Timeline of the observations following the detection of GW170817 with a focus on the high-energy, non-thermal domain. A more complete picture of the multi-wavelength and multi-messenger campaign is given in~\citep{GW170817MMA}.}\label{fig:Timeline}
\end{figure*}%

\subsection{GW170817}
A gravitational wave event was recorded on August 17, 2017, at 12:41:04 UTC by the Advanced LIGO and Advanced Virgo interferometers~\citep{LVCGW170817PRL}. Based on a BAYESTAR analysis using the data of the LIGO-Hanford instrument an initial alert notice was issued at 13:08:16 UTC. A subsequent GCN circular reporting a highly significant detection of a binary neutron star signal was distributed among a wide range of follow-up observatories about 40 minutes after the event at 13:21:42 UTC~\citep{GCN21505}. As only data from a single interferometer was used in this initial reconstruction, the sky location of the event could only be localized to within $24,200\,\mathrm{deg}^2$ ($90\%$ containment). 
Nevertheless, the timing of the alert allowed the team of the \textit{Fermi} Gamma-Ray Burst Monitor (Fermi-GBM) to correlate the GW event with a gamma-ray burst~\citep[170817A,][]{GCN21506, GCN21520} observed $\sim\,1.7\,\mathrm{s}$ after the gravitational-wave candidate. The light-curve of the GRB event shows a weak short pulse with a duration of 2 seconds, typical for \textit{short} GRBs~\citep{GCN21528}. GRB170817A has also been recorded by the SPI-ACS instrument onboard the INTEGRAL satellite~\citep{GCN21507}. Further details are given in~\citet{IntegralGBMLVC}.

\begin{figure*}[!th]
  \resizebox{\hsize}{!}{
  \includegraphics[width=0.98\textwidth]{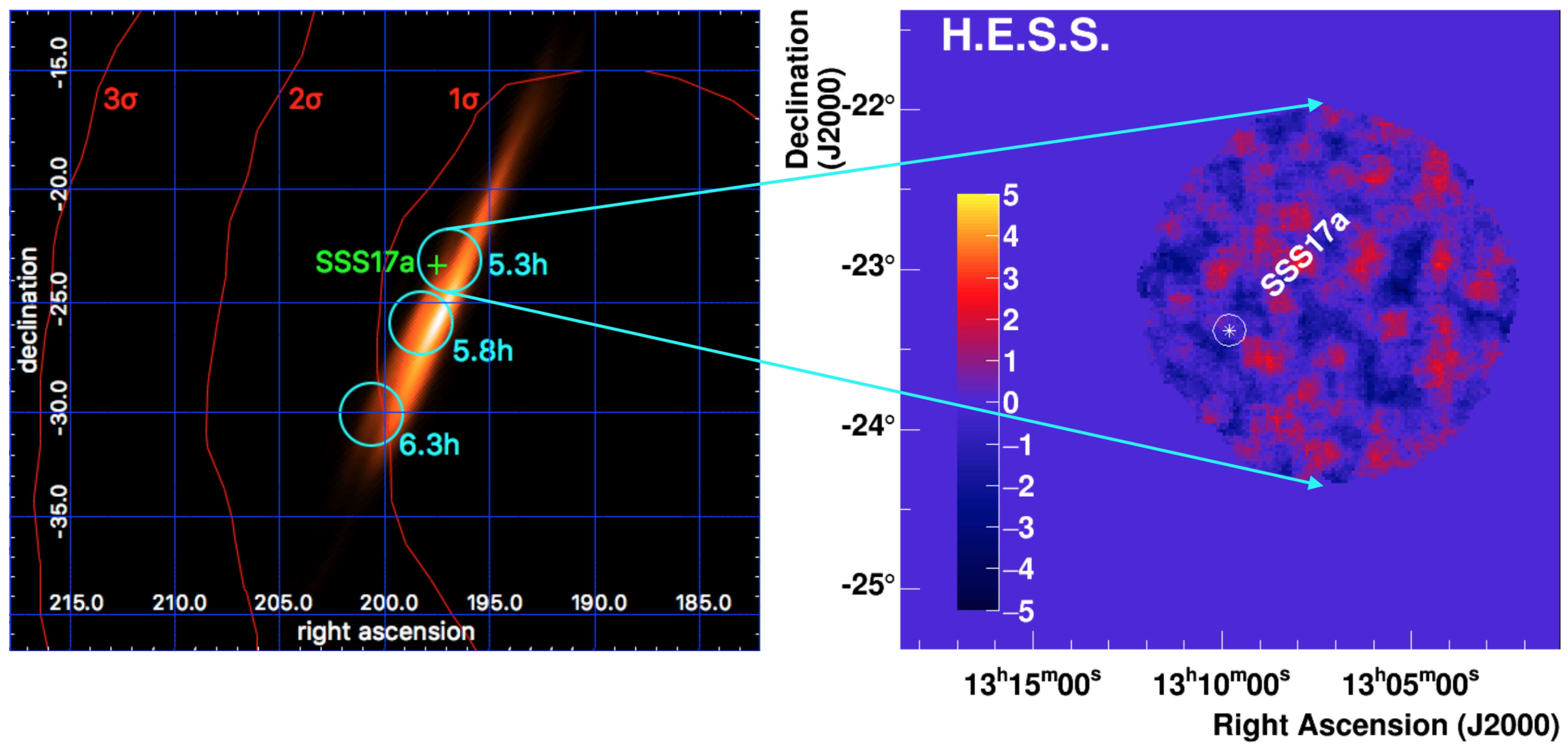}%
  }
\caption{Left: Pointing directions of the first night of H.E.S.S. follow-up observations starting August 17, 2017, at 17:59 UTC. The circles illustrate a FoV with radius of $1.5^\circ$ and the shown times are the starting times of each observation with respect to GW170817. The LALInference map of GW170817 is shown as colored background, the red lines denote the uncertainty contours of GRB170817A. Right: Map of significances of the gamma-ray emission in the region around SSS17a obtained during the first observation of GW170817. The white circle has a diameter of $0.1^\circ$, corresponding to the H.E.S.S. point spread function and also used for the oversampling of the map.}\label{fig:FirstNight}
\end{figure*}

On August 17, 2017, at 17:54:51 UTC, the LIGO-Virgo collaboration provided an update on the gravitational wave skymap, incorporating data from the LIGO Livingston detector (which had to be excluded in the initial analysis due to a noise artifact) as well as data from the Virgo detector in the BAYESTAR pipeline ~\citep[BAYESTAR\_HLV in the following]{GCN21513}. The result of this joint analysis reduced the $90\%$ localization uncertainty of the GW event to about $31\,\mathrm{deg}^2$. The data confirmed the binary neutron star origin and located the merger event at a distance of $40 \pm 8\,\mathrm{Mpc}$ ($50 \pm 3\,\mathrm{Mpc}$ if assuming the binary to be face on). A further analysis using the LALInference method was provided about six hours later ~\citep[2017 August 17, 23:54:40 UTC]{GCN21527}. The $90\%$ credible region of this map (cf. Fig.~\ref{fig:FirstNight}) spans $34\,\mathrm{deg}^2$, overlapping with the $90\,\%$ uncertainty region of GRB170817A~\citep{GRB170817Apaper}. The final estimates of the source properties of GW170817 are given in~\citet{LVCGW170817PRL}.


The first EM counterpart to GW170817 and GRB170817A was detected in the near-infrared by the One-Meter Two-Hemisphere (1M2H) collaboration with the 1\,m Swope telescope at Las Campanas Observatory in Chile on August 17 at 23:33 UTC, i.e. 10.87\,hr after GW170817~\citep{GCN21529}. The source, located at $\alpha(\mathrm{J}2000.0) = 13^{\text{h}}09^{\text{m}}48^{\text{s}}.085\pm0.018$, $\delta(\mathrm{J}2000.0)
= -23^{\circ}22\arcmin53\arcsec.343\pm0.218$, near the early-type galaxy NGC 4993 at a distance of $42.5\,\mathrm{Mpc}$, got designated Swope Supernova Survey 2017a (SSS17a). It had an initial brightness of magnitude $17.3\pm0.1$ in the $i$ band~\citep{GCN21567}. The IAU designation of the source is AT2017gfo. NGC 4993 is on the list of possible candidates that had been identified by "Global Relay of Observatories Watching Transients Happen" network~\citep{GCN21519} via cross-matching the gravitational wave localization with the "census of the local universe" catalogue~\citep{cook2016clu}.  
The optical transient was detected independently by five different teams: the Distance Less Than 40Mpc (DLT40) survey (\citeauthor{GCN21531}), by~\citeauthor{GCN21576},~\citeauthor{GCN21546},~\citeauthor{GCN21530} and~\citeauthor{GCN21538}. Archival searches (e.g. ASAS-SN~\citep{GCN21533}, Hubble~\citep{GCN21536}, etc.) did not show evidence of emission at this position in observations taken before the GW event.
 
The subsequent MWL follow-up campaign focused mainly on the optical transient SSS17a. The monitoring of the source in the UV, optical and near-infrared domain allows the detailed description of its spectral evolution over timescales extending from hours to several days and weeks. The source has also been monitored in UV and X-rays by Swift~\citep{GCN21550_Science} over several days. An X-ray source coincident with the location of SSS17a has been discovered by Chandra about 9 days after GW170817~\citep{GCN21765}.  In the radio domain, the first counterpart consistent with the optical transient position was identified on September 2 and 3, 2017, (16 days after GW170817) by two observations using the Jansky VLA ~\citep{GCN21814, GCN21815}.

This extensive monitoring campaign covering the full EM spectrum, including the high-energy (HE) and very-high-energy (VHE) gamma ray domains (the latter reported in this paper) and searches for high-energy neutrinos, allowed us to monitor the evolution of the source over several days. Focusing on the high-energy, non-thermal domain, a subset of the observations obtained during this campaign is shown in Fig.~\ref{fig:Timeline}. Further details of this unprecedented multi-wavelength and multi-messenger effort can be found in~\citep{GW170817MMA} and references therein.

\subsection{H.E.S.S. follow-up of GW170817}
Here we report on observations obtained in the very-high energy gamma-ray domain with the H.E.S.S. imaging atmospheric Cherenkov telescope array. H.E.S.S. is located on the Khomas Highland plateau of Namibia (23$^{\circ}16'18''$ South, $16^{\circ}30'00''$ East), at an elevation of 1800 m above sea level. With its original four-telescope array, H.E.S.S. is sensitive to cosmic and gamma-rays in the 100\,GeV to 100\,TeV energy range and is capable of detecting a source with an energy spectrum similar to the Crab nebula under good observational conditions close to zenith at the 5$\sigma$ level within less than one minute~\citep{HESS-Crab2006}. In 2012 a fifth telescope with 28\,m diameter was commissioned, extending the covered energy range toward lower energies. The observations reported here were conducted jointly with three of the original 12-m telescopes and the 28-m telescope. One of the 12-m H.E.S.S. telescopes was not available due to a maintenance campaign. 

\subsubsection{Scheduling for GW follow-up}\label{sec:HESSscheduling}
The localization uncertainty derived from the data of the GW interferometers is significant for events detected by two interferometers (hundreds to thousands of square degrees) and still sizable for events with data from three detectors (tens to hundreds of square degrees). Although the field-of-view of the large 28-m H.E.S.S. telescope and the four 12-m telescopes has a radius of about $1.5^\circ$ and $2.5^\circ$, respectively, several pointings are typically necessary to cover the identified region. An additional challenge is related to the limited duty-cycle of the observatory, operating only in astronomical darkness during moonless nights and the accessible range of zenith angles (usually $<60^\circ$). Since H.E.S.S. joined the EM follow-up group of the LIGO-Virgo collaboration, several algorithms have been developed to optimize the follow-up of GW events while taking into account these constrains. The most straightforward and most general scheduling algorithms determine the pointing of the telescopes by maximizing the coverage of the two-dimensional localization probability provided with the GW alerts. In addition to these algorithms, we developed optimized strategies for events occurring at distances for which sufficient complete galaxy catalogs are available. For these we use the GLADE catalogue~\citep{dalya2016vizier}, a value-added full-sky galaxy catalog highly complete and specifically built in order to support EM follow-up of GW signals. It includes more than 3 million entries and is (outside the Galactic plane) complete up to $\sim$ 70 Mpc, well matching the horizon of the current GW interferometers to detect mergers of binary neutron star systems.

Our approach follows the one outlined by~\cite{Goingthedistance}. We use the full three-dimensional information of the location of the GW-event provided by the BAYESTAR and LALInference GW pipelines and correlate it with the location of galaxies within that volume. Several algorithms have been implemented to derive an optimized pointing scenario from this 3D GW-Galaxies probability region. The \textit{One-in-FoV} algorithm searches for the coordinates that provide the highest probability of hosting the event, while the \textit{Gal-in-FoV} algorithm determines the center of a region on the sky which provides best coverage of neighboring high-probability regions falling in the same Field of View (FoV). Both algorithms are taking into account observational constraints like the available time window and, trying to achieve a low energy threshold, optimize the pointing strategy favoring low-zenith angle observations. Both are complementary in terms of calculation speed and performance, with \textit{One-in-FoV} being used for real-time follow-ups and the \textit{Gal-in-FoV} for offline scheduling. Further details about the developed approaches and performance estimates based on Monte Carlo simulations of NS-NS merger events are given in~\cite{MoriondVHEPU2017}. 

\subsubsection{Scheduling for GW170817}\label{sec:HESSschedulingGW170817}
As outlined above, the first localization map for the event GW170817 was provided by the BAYESTAR pipeline and was made available to follow-up partners about $1.5\,\mathrm{h}$ after the GW event~\citep{GCN21509}. Due to its large uncertainty covering $24,200\,\mathrm{deg}^2$ at $90\%$ containment, it was not suitable for scheduling follow-up observations. An updated BAYESTAR-reconstructed GW map, BAYESTAR\_HLV, using data from all three interferometers was received about 5 hours after the event, at 17:54 UTC~\citep{GCN21513}. This map, with the $90\%$ region of the localization uncertainty covering $31\,\mathrm{deg}^2$, was used for the scheduling of H.E.S.S. follow-up observations. With H.E.S.S. data taking starting on August 17 at 17:59 UTC, only about 5 minutes were available to derive a pointing strategy. A LALInference based skymap was made available about 9 hours after the gravitational wave event. Changes with respect to the low-latency BAYESTAR\_HLV map were minimal (the $90\%$ uncertainty region increased slightly to $34\,\mathrm{deg}^2$). 

Due to the limited time between the publication of the BAYESTAR\_HLV map and the start of the visibility window we used the \textit{One-in-FoV} approach to determine the H.E.S.S. pointing schedule for the night of August 17/18. Three observation runs of $28\,\mathrm{min}$ each were scheduled. They are given in Tab.~\ref{tab:HESSPointings} and illustrated in Fig.~\ref{fig:FirstNight}. The three observations were taken between August 17, 17:59 - 19:30 UTC. For an effective FoV of $1.5^\circ$ radius of the H.E.S.S. 28\,m telescope, they cover about $56\,\%$ of the gravitational wave uncertainty region of the final LALInference map. At the same time they include about 86\% of the probability density region obtained by weighting the three-dimensional GW map with galaxies from the GLADE catalog. All three pointings are compatible with GRB1707A within 2$\sigma$. Whereas the optical transient SSS17a had not been discovered at the time our observations took place, the focus on a region containing many galaxies compatible with the 3D-GW map allowed us to cover NGC 4993 and SSS17a with our first observation, i.e. starting 5.3h after the GW event. We note that our observations have the shortest time delay with respect to GW170817 by any ground-based pointing instrument participating in the follow-up of GW170817.

After the detection of SSS17a during the night of August 17/18 we discontinued further coverage of the GW uncertainty region and focussed on monitoring the source in TeV gamma-rays. H.E.S.S. observations were scheduled at the beginning of the following nights around SSS17a as long as the location was visible from the H.E.S.S. site within a maximum zenith angle of about $60^\circ$ and fulfilling the necessary observation conditions. The obtained observations are summarized in Table \ref{tab:HESSPointings}. 

\begin{table}[!ht]
\caption{H.E.S.S. follow-up observations of GW170817. All pointings were taken with the default run duration of $28\,\mathrm{min}$ and are given in equatorial J2000 coordinates.}
\label{tab:HESSPointings}
\begin{tabular}{lcccc}
ID & Observation time  & Pointing coordinates  & <zenith angle> \\
 & (UTC)& [deg]& [deg] \\
\hline
1a & 2017-08-17   17:59  &  196.88,  -23.17  &  59  \\
1b & 2017-08-17   18:27  &  198.19,  -25.98  &  58  \\
1c & 2017-08-17   18:56  &  200.57,  -30.15  &  62  \\
2a & 2017-08-18   17:55  &  197.75,  -23.31  &  53  \\
2b & 2017-08-18   18:24  &  197.23,  -23.79  &  60 \\
3a & 2017-08-19   17:56  &  197.21,  -23.20  &  55  \\
3b & 2017-08-19   18:24  &  197.71,  -23.71  &  60  \\
5a & 2017-08-21   18:15  &  197.24,  -24.07  &  60 \\
6a & 2017-08-22   18:10  &  197.70,  -24.38  &  60  \\
\end{tabular}
\end{table}

\section{Data and analysis}\label{Section:Data}
\begin{figure*}[!thp]
 \centering
 \begin{tabular}[b]{@{}p{0.47\textwidth}@{}}
   \centering\includegraphics[width=1.05\linewidth]{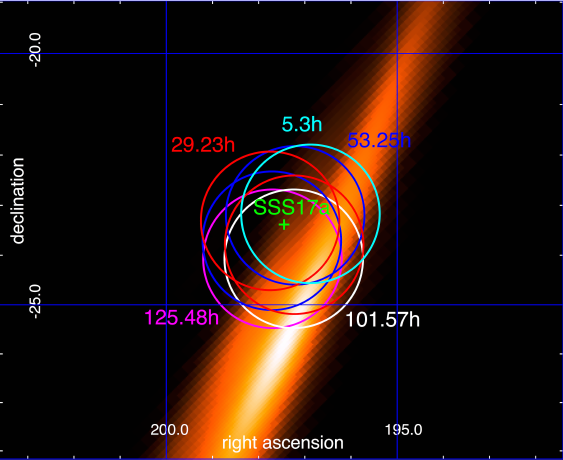} \\
   \centering\vspace{5.5mm} \small (a) SSS17a: H.E.S.S. pointings
 \end{tabular}%
 \hfill
 \begin{tabular}[b]{@{}p{0.47\textwidth}@{}}
 \vspace{-7.35cm}%
   \centering\includegraphics[width=1.0\linewidth]{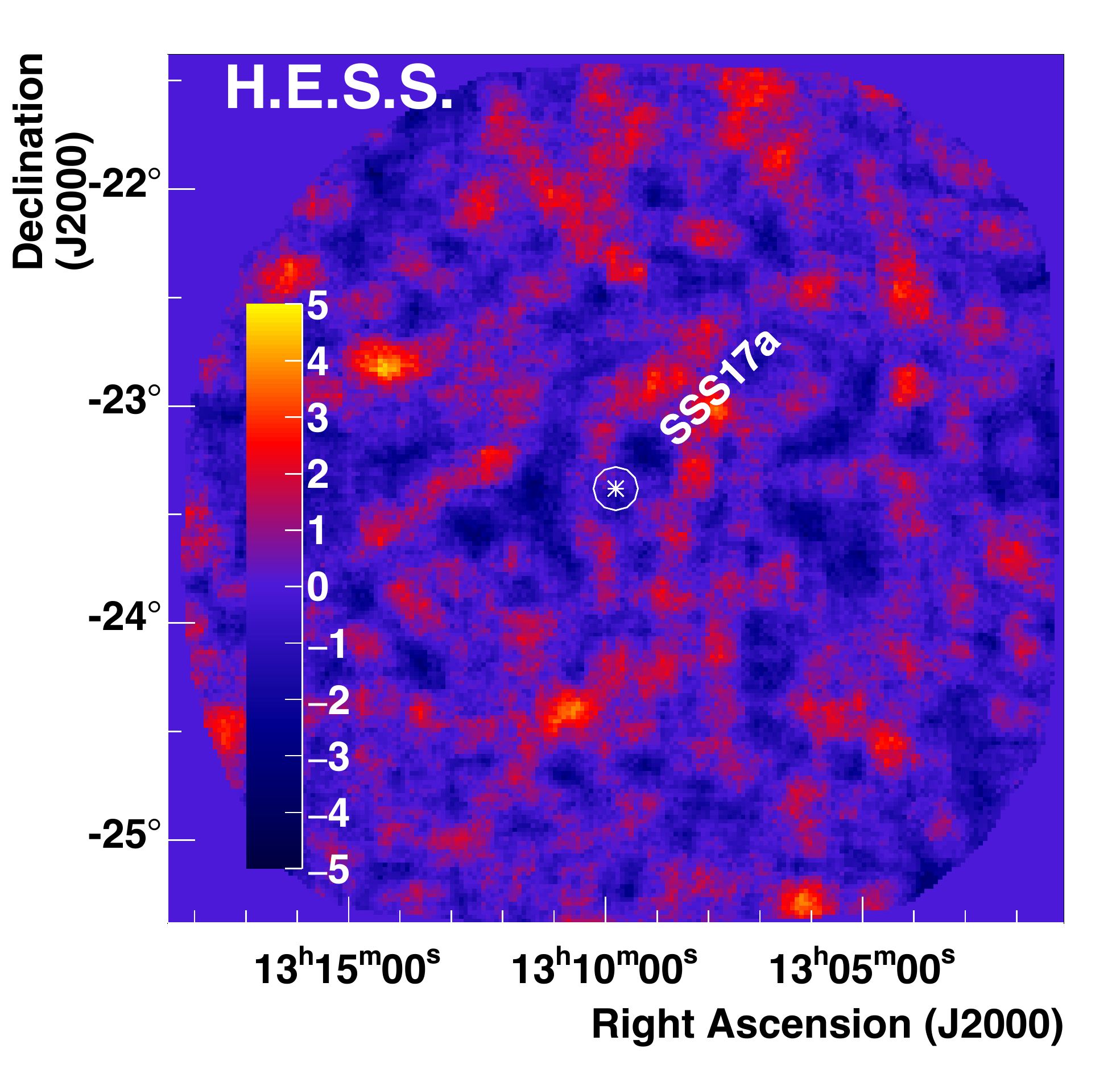} \\
   \centering\vspace{-1mm}\small (b) SSS17a: H.E.S.S. significance map
 \end{tabular}
 \caption{Left plot: H.E.S.S. pointing directions during the monitoring campaign of SSS17a. For details see Tab.~\ref{tab:HESSPointings}. The circles denote a FoV with radius of $1.5^\circ$ and the shown times are the start times of each observation with respect to GW170817. Right plot: Map of significances of the gamma-ray emission in the region around SSS17a combining all observations obtained during the H.E.S.S. monitoring campaign.}\label{fig:SSS17aPointings}
\end{figure*}%
The obtained data were analyzed using Model Analysis~\citep{ModelAnalysis}, an advanced Cherenkov image reconstruction method in which the recorded shower images are compared to a semi-analytical model of gamma-ray showers by means of a log-likelihood optimization. The background level in the FoV was determined from the dataset itself using the standard ``ring background'' technique~\citep{RingBg}. Relying on the azimuthal symmetry of the response of the telescopes, the required acceptance function has been derived from the data itself.
We perform our analysis using only data from the 28\,m telescope in the center of the H.E.S.S. array in order to achieve a low energy threshold. We adopted the ``Loose cuts'' of the Model Analysis which for example require the total charge in the recorded shower image to be greater than 60 photoelectrons. This and additional quality selection criteria yield an energy threshold of 280\,GeV for the first observation and 270\,GeV for the combined dataset on SSS17a. We note that the threshold is significantly influenced by the relatively high zenith angle of the observations.  We further require that at least 10 events are available for the background estimation, a requirement that limits the energy range over which our results are valid. The derived energy ranges are given in Tab.~\ref{tab:limits}. Further analyses exploiting the data from the full H.E.S.S. array will be published at a later time.

A second analysis using a fully independent data calibration chain and the Image Pixel-wise fit for Atmospheric Cherenkov Telescopes~\citep[ImPACT,][]{ImPACT} reconstruction method was used to verify the results. The results of this cross-check analysis are consistent with the ones presented here, thus providing confidence in the robustness of the presented results.

High-energy gamma rays interact with the extragalactic background light (EBL) via $e^+ / e^-$ pair-creation processes. At the highest energies gamma rays are thus absorbed during the propagation through the extragalactic radiation fields. The resulting opacity depends on the gamma-ray energy and the distance of their source. We used the EBL model published in~\citet{Franceschini2008} to calculate these energy dependent EBL correction factors. Using the redshift of NGC4993, $z=0.009787$~\citep{2000A&AS..143....9W}, these factors increase with energy and are about $10\,\%$ ($30\,\%$) at $1\,\mathrm{TeV}$ ($10\,\mathrm{TeV})$. These effects are therefore only of minor importance and we here do not correct for them in this paper.

The region covered by our observations contains several sources with emission in the GeV energy range. They are for example listed in the catalog compiled after four-year long observations by the LAT instrument onboard the Fermi satellite~\citep{3FGL}. None of them is known to exhibit emission in the TeV range~\footnote{\tt http://tevcat.uchicago.edu}. The most promising TeV candidate source in the region is likely PKS 1309-216, at an angular distance of $1.58^\circ$ from NGC4993/SSS17a. It has a flux in the $1-100\,\mathrm{GeV}$ range of about $2.0\times10^{-11}\,\mathrm{erg\,cm^{-2}\,s{-1}}$ and an energy spectrum following $E^{-2.07 \pm 0.05}$ in the same energy range. 
Extrapolating the emission to the higher energies relevant for our observations needs to account for EBL absorption effects: at the redshift of PKS 1309-216 ($z=1.489$~\citep{2000A&AS..143....9W}) the initial flux is decreased by more than one order of magnitude at energies around $100\,\mathrm{GeV}$ and by a factor exceeding $10^{9}$ at $1\,\mathrm{TeV}$~\citep{Franceschini2008}. Conservatively, we nevertheless excluded a region with a $0.3^\circ$ radius around PKS 1309-216 from the background estimation used for the analysis presented here.

None of the GeV detected sources showed significant flux increases during the period of the observations presented here~\citep{2013ApJ...771...57A}. We therefore conclude that no TeV gamma-ray emission exceeding the level of the reached sensitivity, other than a potential signal related to GW170817 and GRB170817A, is expected.

\begin{figure*}[!thp]
 \centering
 \begin{tabular}[b]{@{}p{0.47\textwidth}@{}}
   \centering\includegraphics[width=0.98\linewidth]{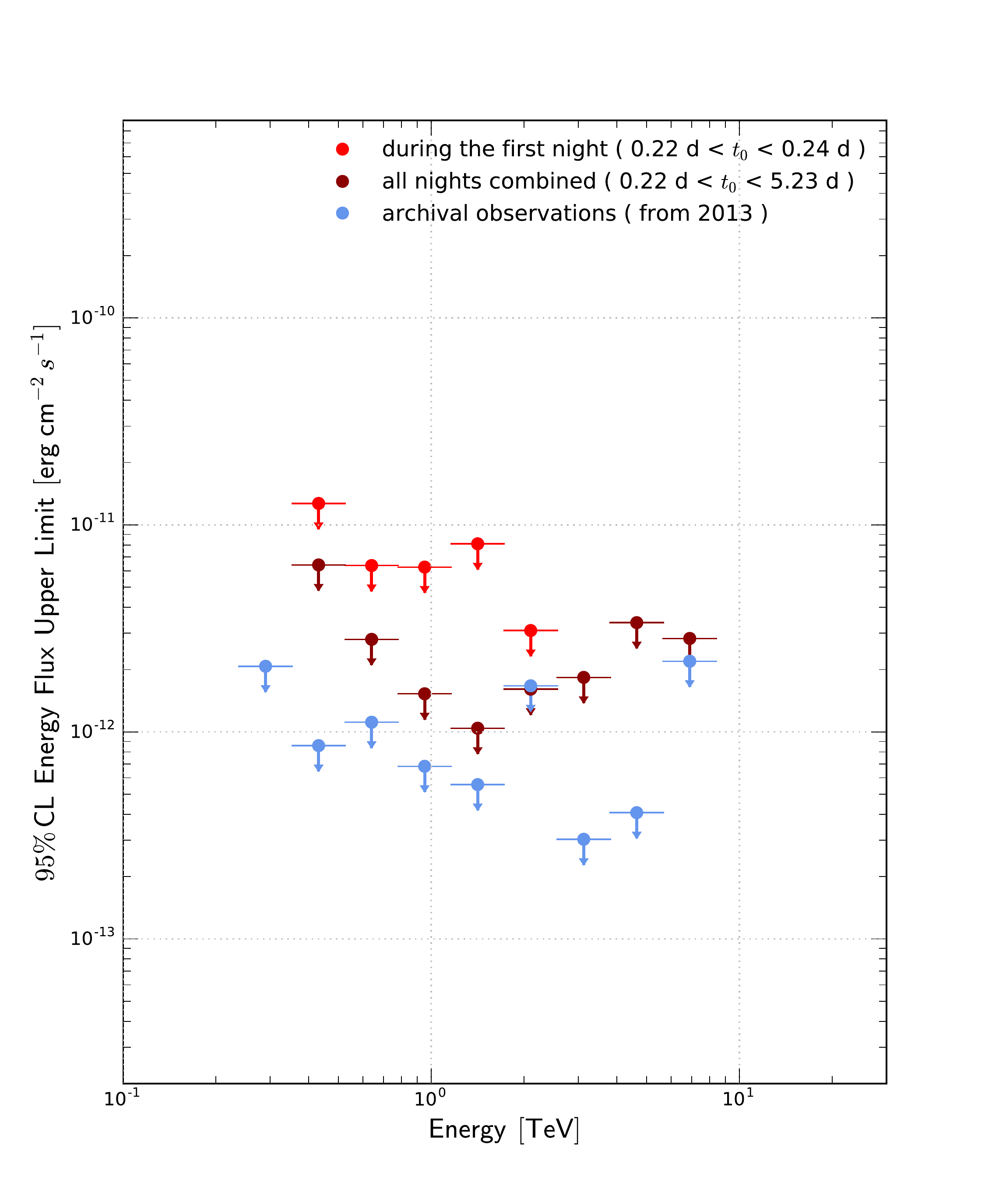} \\
   \centering\small (a) SSS17a: H.E.S.S. limits
 \end{tabular}%
 \hfill
 \begin{tabular}[b]{@{}p{0.47\textwidth}@{}}
   \vspace*{-9.8cm}\centering\includegraphics[width=1.04\linewidth]{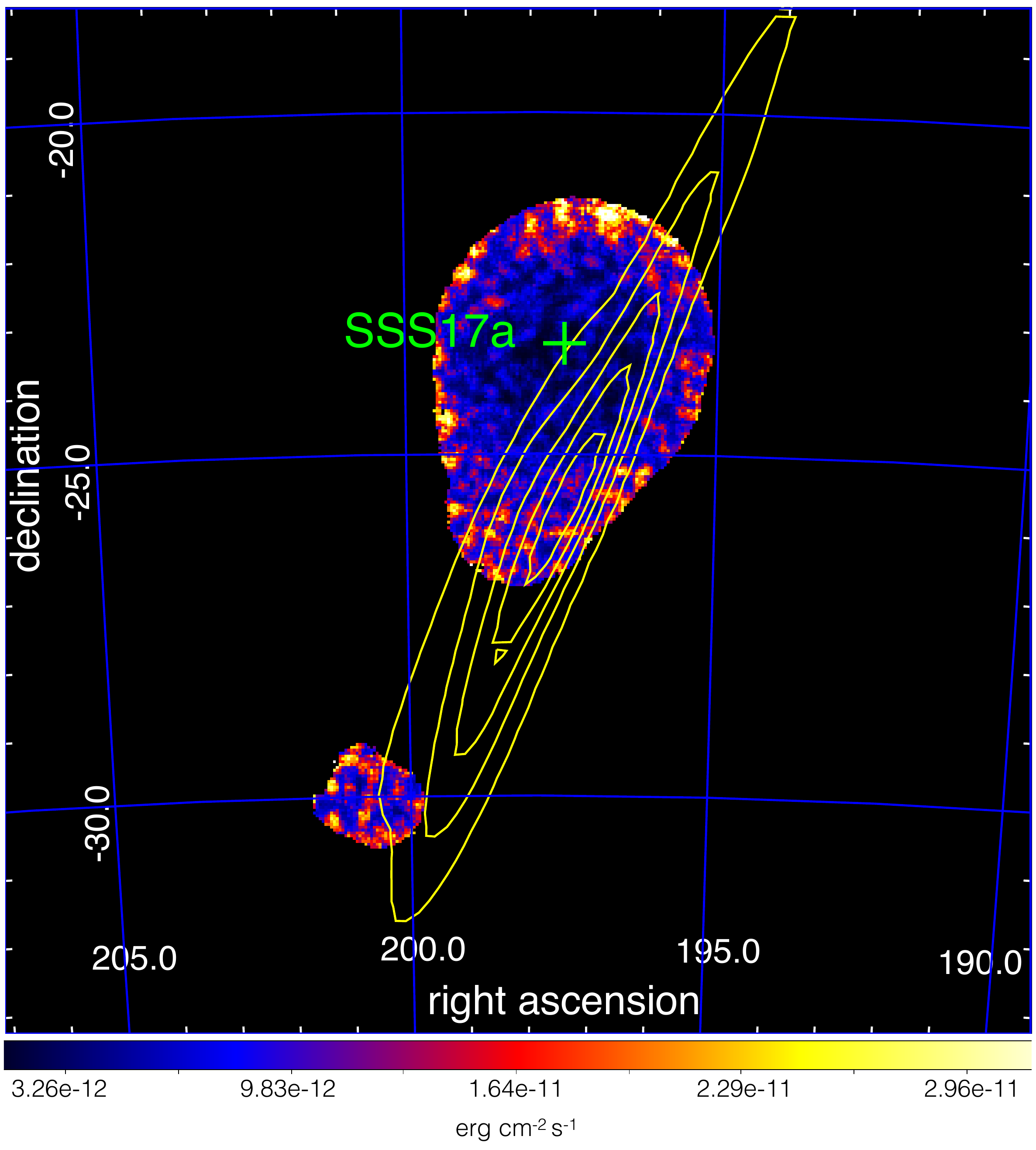} \\
   \centering\small (b) GW170817: H.E.S.S. flux limit map
 \end{tabular}
 \caption{Left plot: Differential upper limits on the gamma-ray flux from SSS17a derived from the H.E.S.S monitoring campaign and archival observations of the region. Right plot: Map showing the integral upper limits in the $270\,\mathrm{GeV}$ to $8.55\,\mathrm{TeV}$ energy range (assuming an $E^{-2}$ energy spectrum) derived from the H.E.S.S. follow-up observations of GW170817. The yellow contours outline the localization of the GW event as provided by the LALInference map.}\label{fig:HESSall}
\end{figure*}

We note that archival H.E.S.S. observations on PKS 1309-216 have been obtained in 2013. After about $10\,\mathrm{h}$ of observations, neither gamma-ray emission from the source nor from the region around NGC 4993 / SSS17a could be detected. We therefore derive an archival upper limit on the gamma-ray flux at from SSS17a from these observations to $\Phi(170\,\mathrm{GeV}\,<\,E\,<\,47.2\,\mathrm{TeV}) < 3.1\times 10^{-12}\,\mathrm{erg\,cm^{-2}\,s^{-1}}$ at $95\%\,\mathrm{C.L.}$ and assuming a spectral index of -2. The differential upper limit as function of energy is shown in the left plot of Fig.~\ref{fig:HESSall}.

\section{Results}\label{Section:Results}
As outlined above, our observations of SSS17a started August 17, 2017, at 17:59 UTC (pointing 1a) and were repeated during several nights (cf. Tab.~\ref{tab:HESSPointings}). The different pointings, except 1b and 1c, which are not covering SSS17a but were taken during the initial scanning of the GW170817 uncertainty region, are shown in the left plot of Fig.~\ref{fig:SSS17aPointings}. The same color is used for pointings obtained successively during the same night and the numbers indicate the time difference between the start of the observations and the time of GW170817. As the potential gamma-ray emission from a NS-NS merger is expected to be transient, we analyzed each of the obtained nightly datasets independently. For each of them we produce skymaps of the gamma-ray excess counts above the background derived from the data itself as described above. These excess maps have then been converted into significance maps using the formalism described by~\citet{LiMa}. 

As example, we show the gamma-ray significance map derived from the first observation in Fig.~\ref{fig:FirstNight}. An oversampling radius of $0.1^\circ$, roughly corresponding to the H.E.S.S. point-spread function has been applied. No significant gamma-ray emission is found within any of the individual datasets and all of the obtained results are fully compatible with the background-only expectation. We thus conclude that no significant VHE gamma-ray afterglow was detected from the direction of SSS17a. Consequently we derive $95\,\%$ C.L. upper limits on the gamma-ray flux following~\citet{FeldmanCousins}. The obtained flux limits, assuming a generic $E^{-2}$ energy spectrum for the potential emission, are given together with the corresponding energy ranges in Tab.~\ref{tab:limits} and shown in Fig.~\ref{fig:Timeline}.

\begin{table}[htp]
\caption{Limits on the high-energy gamma-ray flux at $95\,\%$ C.L. and assuming a $E^{-2}$ energy spectrum obtained during the monitoring of SSS17a with H.E.S.S.}\label{tab:limits}
\begin{center}
\begin{tabular}{cccc} 
pointings  &  time since GW& $f_\gamma$   & Energy band \\
 (see Tab.\ref{tab:HESSPointings}) & trigger [days] & [$\mathrm{erg\,cm^{-2}\,s^{-1}}$]  &[TeV]  \\
 \hline
 1a & 0.22 &  $<3.9\times 10^{-12}$ & 0.28--2.31   \\
2a+2b & 1.22 & $<3.3\times 10^{-12}$  & 0.27--3.27   \\
3a+3b  & 2.22 & $<1.0\times 10^{-12}$ & 0.31--2.88  \\
5a+6a & 4.23,\,5.23 & $<2.9\times 10^{-12}$ & 0.50--5.96  \\
all & 0.22 -- 5.23 & $<1.5\times 10^{-12}$ & 0.27--8.55 
\end{tabular}
\end{center}
\label{default}
\end{table}%

In a search for fainter but temporally extended emission from SSS17a, we combined all datasets (except 1b and 1c). The obtained significance map (cf. right plot of Fig.~\ref{fig:SSS17aPointings}) is again fully compatible with the background-only hypothesis. We obtain $\Phi_\gamma < 1.5\times 10^{-12}\,\mathrm{erg\,cm^{-2}\,s^{-1}}$ in the energy band $0.27 < E [\mathrm{TeV}] < 8.55$. Assuming a radially symmetric emission this flux limit corresponds to a limit on the VHE gamma-ray luminosity of SSS17a at a distance of $42.5\,\mathrm{Mpc}$ of $L_\gamma < 3.2\times 10^{41}\,\mathrm{erg\,s^{-1}}$. We note the luminosity of the prompt phase of GRB170817A that has been found to be around $2.2\,\times\,10^{46}\,\mathrm{erg\,s^{-1}}$ by INTEGRAL SPI-ACS~\citep{GCN21507}. Differential upper limits as function of the energy are shown in the left plot of Fig.~\ref{fig:HESSall} for the first observation on SSS17a, the combined dataset and the archival observations obtained in 2013.

After combining all observations obtained with H.E.S.S. during the follow-up campaign of GW170817 we derive a skymap showing the integral upper limits in the $270\,\mathrm{GeV}$ to $8.55\,\mathrm{TeV}$ energy range. It is shown in the right plot of Fig.~\ref{fig:HESSall}. First of all, it illustrates the deep observations centered on SSS17a. Induced by the radially decreasing acceptance of the telescope, the obtained limits are less constraining when approaching the border of the field-of-view. The figure also illustrates the achieved $\approx 50\%$ coverage of the LALInference map of GW170817, which is depicted by the yellow contours.

\section{Discussion and conclusion}\label{Section:Summary}
The observations presented here represent the first very-high-energy gamma-ray observations following the merger of a binary neutron star system. A preprepared scheduling procedure allowed fast reaction to the event and provided efficient pointings within the gravitational wave uncertainty region, covering observational fields including that of the multi-wavelength counterpart SSS17a even before it had been discovered from optical observations. Following the discovery of this counterpart in the optical band, subsequent extended monitoring allowed deep observations to be made of this source. Although the source was not detected within the energy range $0.27 < E [\mathrm{TeV}] < 8.55$, the derived upper limits are the most stringent ones obtained on hour to week-long timescales, of non-thermal emission from GW170817 in the full gamma-ray domain ranging from keV to TeV energies. They allow for the first time a constraint to be placed on the level of early-time very-high-energy emission from the source, following the binary neutron star merger. With a potential connection to a {\it kilonova} type event, expected to give rise to the ejection of mildly relativistic outflows, further observations of this object should be performed to probe particle acceleration beyond TeV energies on longer timescales. 

\begin{acknowledgements}
The support of the Namibian authorities and of the University of Namibia in facilitating the construction and operation of H.E.S.S. is gratefully acknowledged, as is the support by the German Ministry for Education and Research (BMBF), the Max Planck Society, the German Research Foundation (DFG), the French Ministry for Research, the CNRS-IN2P3 and the Astroparticle Interdisciplinary Programme of the CNRS, the U.K. Science and Technology Facilities Council (STFC), the IPNP of the Charles University, the Czech Science Foundation, the Polish Ministry of Science and Higher Education, the South African Department of Science and Technology and National Research Foundation, the University of Namibia, the Innsbruck University, the Austrian Science Fund (FWF), and the Austrian Federal Ministry for Science, Research and Economy, and by the University of Adelaide and the Australian Research Council. We appreciate the excellent work of the technical support staff in Berlin, Durham, Hamburg, Heidelberg, Palaiseau, Paris, Saclay, and in Namibia in the construction and operation of the equipment. This work benefited from services provided by the H.E.S.S. Virtual Organisation, supported by the national resource providers of the EGI Federation.
\end{acknowledgements}

\end{document}